\documentclass[12pt]{article}

\usepackage{sbc-template}

\usepackage{graphicx,url}

\usepackage[british,UKenglish,USenglish,english,american]{babel}
\usepackage[utf8]{inputenc}  
\usepackage{amsmath}
\DeclareMathOperator{\grad}{grad}
\DeclareMathOperator{\divv}{div}
 
\usepackage{lipsum} %for footnotes 
\newcommand\blfootnote[1]{%
  \begingroup
  \renewcommand\thefootnote{}\footnote{#1}%
  \addtocounter{footnote}{-1}%
  \endgroup
}
     
\title{Dynamics of fracturing saturated porous media and self-organization of rupture}

\author{C. Peruzzo\inst{1}, D.T. Cao\inst{2}, E. Milanese
  \inst{3}, P. Favia\inst{1}, F. Pesavento\inst{1}, F. Hussain\inst{2},\\ B.A. Schrefler\inst{1} }

\address{University of Padova, Dept. of Civil, Environmental and Architectural Engineering, Italy
\nextinstitute
  Texas Tech University, Dept. of Mechanical Engineering, USA 
  Durham, U.K.
\nextinstitute
  EPFL (Ecole Polytechnique Federale de Lausanne),\\
   Computational Solid Mechanics Laboratory LSMS, Switzerland
}
\begin{document} 
\maketitle
\blfootnote{\\
The paper was presented as a general lecture at the EUROMECH Solid Mechanics Conference in Bologna.\\
E-mail addresses: mailcarloperuzzo@gmail.com (C.Peruzzo), toancaoduc@gmail.com (D.T.Cao), enrico.milanese@epfl.ch (E.Milanese), pietro.favia86@gmail.com (P.Favia), francesco.pesavento@dicea.unipd.it (F.Pesavento), fazlehussain@gmail.com (F.Hussain), bernhard.schrefler@dicea.unipd.it (B.A.Schrefler).}

\begin{abstract}
  Analytical solutions and a vast majority of numerical ones for fracture propagation in saturated porous media yield smooth behavior while experiments, field observations and a few numerical solutions reveal stepwise crack advancement and pressure oscillations. To explain this fact, we invoke self-organization of rupture observed in fracturing solids, both dry and fully saturated, when two requirements are satisfied: i) the external drive has a much slower timescale than fracture propagation; and ii) the increment of the external load (drive) is applied only when the internal rearrangement of fracture is over. These requirements are needed to obtain clean Self Organised Criticality (SOC) in quasi-static situations. They imply that there should be no restriction on the fracture velocity i.e. algorithmically the fracture advancement rule should always be independent of the crack velocity. Generally, this is not the case when smooth answers are obtained which are often unphysical. Under the above conditions hints of Self Organized Criticality are evident in heterogeneous porous media in quasi-static conditions using a lattice model, showing stepwise advancement of the fracture and pressure oscillations. We extend this model to incorporate inertia forces and show that this behavior still holds. By incorporating the above requirements in numerical fracture advancement algorithms for cohesive fracture in saturated porous continua we also reproduce stepwise advancements and pressure oscillations both in quasi-static and dynamic situations. Since dynamic tests of dry specimens show that the fracture advancement velocity is not constant we replicate such an effect with a model of a debonding beam on elastic foundation. This is the first step before introducing the interaction with a fluid.
\end{abstract}

\section{Introduction}
The term “self-organization of rupture” was first coined by Tzschicholz and Herrmann [1] when investigating pressure fluctuations and acoustic emission in hydraulic fracturing. According to Herrmann [2] it expresses an experimentally observed fact in disordered media where systems of cracks organize themselves to create a complex structure (often self-similar). Examples include localization of shear bands out of micro-cracks, formation of a drying cracks network, fractal cracks in hydraulic fracture or stress corrosion cracking and tectonic fault systems. By simulating hydraulic fracturing on a beam lattice model Tzschicholz and Herrmann found that the sequence of breaking events is organized in bursts, also called avalanches, that have power law distributed lifetime and quiescence intervals [3-10]. This indicates that self-organized criticality (SOC) [1] takes place or at least implied for samples of finite dimensions. Ignoring the requirements for self-organized criticality has serious consequences when designing the fracture advancement rule for analytical or numerical simulation of fracturing saturated homogeneous porous media - both in quasi-static and dynamic situations.\\
\newline
Fracturing in fully saturated porous media is of great importance in geophysics [11-15] and in non-conventional oil and gas extraction [16-18]. In geophysics both mechanical [12] action and pressure induced fracture [11, 19] are the case while in oil and gas extraction the latter is prevailing. A signature of hydraulic fracturing through lava is e.g. the magmatic dikes radiating from West Spanish Peak, Colorado, U.S. [20]. Despite the fact that in pressure induced fracture in geological situations fracture advancement speed can reach 1/5th of the speed of sound [11] and slugs of magma can rise in rocks up to a speed of 17m/s [21], inertia forces are usually neglected in simulations. Some solutions including inertia forces exist, however [22-25]. [22,23] show continuous fracture growth, [24] some pressure oscillations due to fracture toughness contrast between two adjacent layers, while [25] clearly exhibits stepwise behavior and pressure oscillations. This aspect will be addressed below and involves the use of algorithms which respect “self-organization of rupture”.\\
\newline
The many existing closed form and numerical solutions, see [25, 26] for extensive lists, are obtained mainly in a quasi-static context and a homogeneous setting. Stochastically heterogeneous fracturing media have been taken into account via statistical analysis in [1] as already mentioned, but without flow in the domain. Flow in the fracture and in the domain, has been investigated in [27] on a central force model for mechanical load and pressure induced fracture. Also, in these cases the fracture does not propagate in a continuous manner, but advances stepwise and the avalanches size (number of failing elements per loading step) follows a power law. However, it has also been found in [27] that in pressure induced fracture a power law exists only for low injection rates, while for higher injection rates the power law behavior breaks down, see Figure 1.\\
\newline
\begin{figure}[ht]
\centering
\includegraphics[width=1\textwidth]{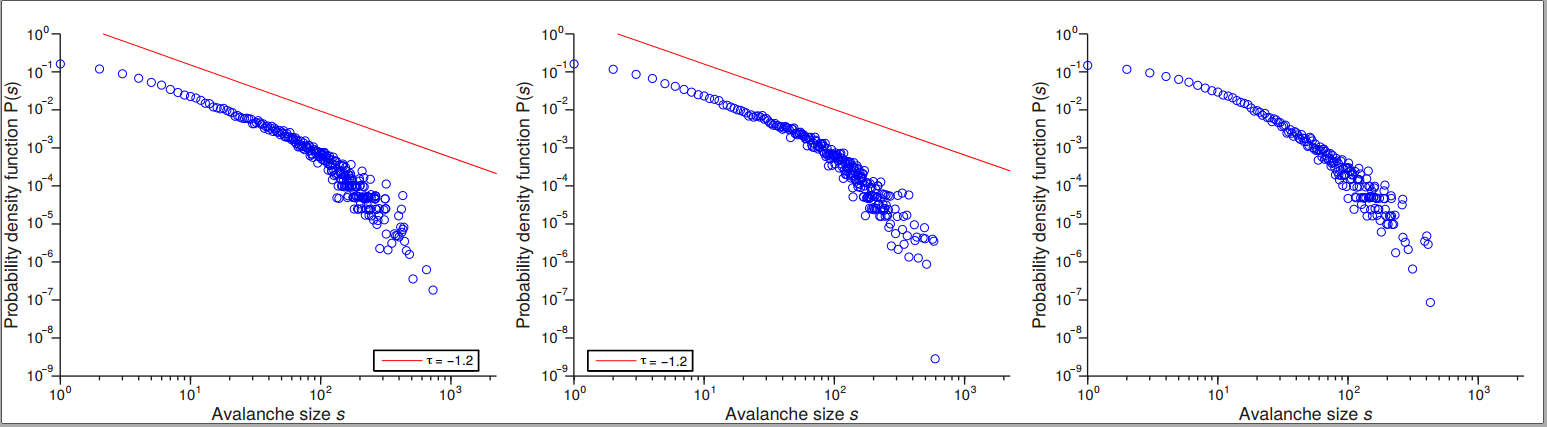}
\caption{Power-law behavior of the avalanche size probability distribution P(s) is displayed for events that take place in the plastic plateau for injection rates of $\boldsymbol{f_{p,inc}=10^{-5} mm^3/s}$  and $\boldsymbol{10^{-4} mm^3/s}$ (left and center);  for an injection rate of $\boldsymbol{f_{p,inc}=10^{-3} mm^3/s}$  the power-law behavior is destroyed (right). Analyses were performed on systems of size L = 16.}
\end{figure}
This hints at a transition in the behavior regime depending on the injection rate, which requires further research. \\
\newline
The stepwise fracture growth and ensuing pressure oscillations are well known in the oil and gas industry since the late 80ties [17,18], Figure 2. There it has important implications for the hydrocarbon production [28]. An experiment of fluid injection in Colton Sandstone [29] evidenced that for high viscosity fluid and low constant injection rate a continuous fracture propagation was obtained (see for example Fig. 3) while high constant injection rate and low viscosity (field conditions for fracking) resulted in a stepwise advancement with ensuing pressure oscillations, Figure 2. Again, there is a transition in behavior depending on the injection rate as well as the viscosity. In [25] we conjectured that the transition from continuous to staccato advancement hides a stability problem, which has still to be proved. \\
Stepwise advancement of a mechanically loaded hydrogel, both in liquid and in air has also been evidenced experimentally [30]. The stepwise behavior was mainly ignored in the mechanics and computational mechanics community. In this community, a first solution evidencing staccato fracture advancement was shown in a quasi-static setting with a standard Galerkin Finite element method with remeshing [31]. Only very recently other solutions capturing the stepwise advancement appeared; in the most recent papers [32,33] the Extended Finite Element Method is used.\\
\begin{figure}[ht]
\centering
\includegraphics[width=0.7\textwidth]{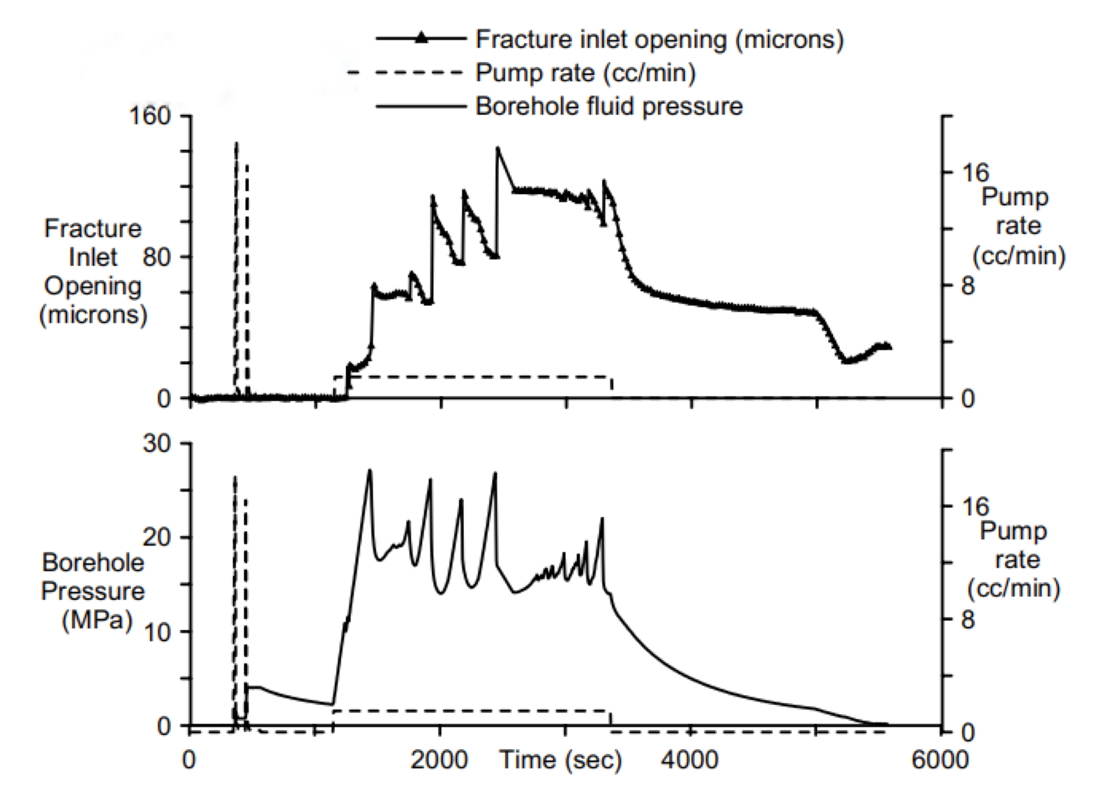}
\caption{Experimental fracture inlet opening and pressure record for low viscosity, high flow rate configuration; reproduced with permission from [20].}
\end{figure}
\begin{figure}[ht]
\centering
\includegraphics[width=0.7\textwidth]{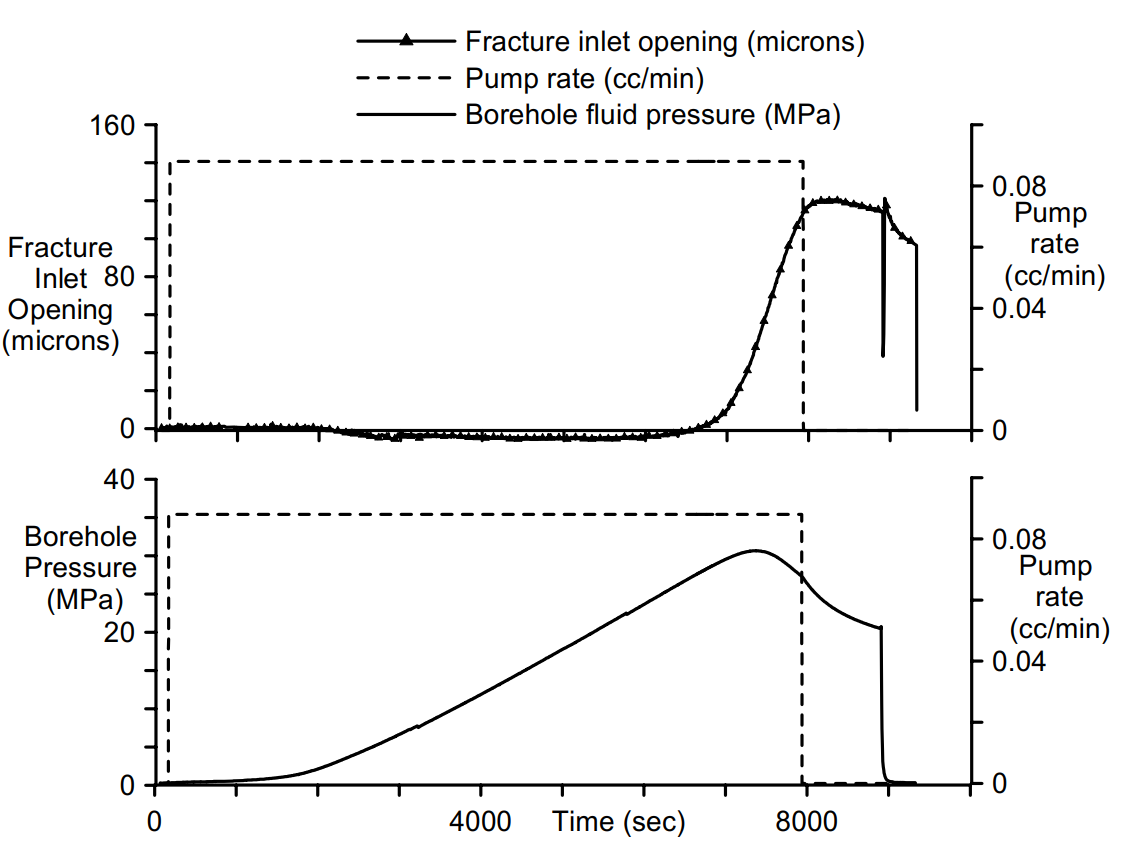}
\caption{experimental pressure record and fracture inlet opening for high viscosity and low flow rate configuration; reproduced with permission from [21].}
\end{figure}
We concentrate here our attention on the solid part and the interaction with the fluid. The behavior of the fluid, especially in the fracture will require further scrutiny. After addressing the issue of self-organization of rupture and the ensuing consequences for the design of the fracture advancement rule we shall extend the central force model of [27] to also include inertia forces and shall show solutions for quasi-static and dynamic analyses in stochastically heterogeneous saturated porous media. We shall consider both mechanical and hydraulic loading. Since it has been shown [5] that inclusion of inertia forces does not drive the system away from criticality, in the dynamic case we will omit the investigation of avalanche behavior, i. e. we shall limit us to single runs. Then we will show solutions obtained with the standard Galerkin Finite Element Method for fracturing homogeneous media in dynamics, based on an algorithm respecting the requirements for self-organization of rupture. Finally, we show a closed form solution for dry materials dynamics, based on a cohesive model. This is a first step towards investigating the above-mentioned conjecture. Concluding remarks will close the paper.

\section{Self-organization of rupture} \label{sec:firstpage}

When studying fracturing of heterogeneous media with and without fluid with methods of statistical physics in quasi-static situations, the following two requirements have to be satisfied, apart from heterogeneity [1, 3, 27]\\
\begin{enumerate}
\item External drive has a much slower timescale than fracture propagation;
\item increments of the external load (drive) are applied only when the internal rearrangement of fracture is over.
\end{enumerate}
By satisfying these conditions, power law behavior has been found for the avalanche size probability distribution (breaking or damaging events in a time step), distribution of time duration of each avalanche, energy bursts, the probability distribution of finding two breaking events to be $\tau$ time steps apart. This last one corresponds to Omori's well-known law of aftershocks for earthquakes first formulated in 1894 [34] which has been verified from earthquake catalogues for aftershocks series ranging from a few hours to a couple of years after the main event [35]. Tzschichholz and Herrmann [1] conclude that one might expect the basic mechanism for burst sequences, or, respectively, aftershocks, to be universal due to self-organization of rupture. 
Also, faster loading than required above is possible, but then we may not be able to discern individual avalanches anymore. Clearly the avalanche behavior of the breaking events entails pressure oscillations with irregular intervals and amplitudes, which persist also in homogeneous media [1,36]. Satisfying condition i) and ii) implies that there should be no restriction on the fracture velocity i.e. algorithmically the fracture advancement rule should always be independent of the crack velocity. This is completely ignored when imposing velocity restriction to obtain solutions. Such velocity restrictions are imposed for instance when searching for analytical solutions, or with inadequate crack tip advancement/time stepping algorithms in numerical solutions. For the first case the statement “the solution near the tip of a propagating finite hydraulic fracture is captured by a stationary solution for a semi-finite crack moving at a constant velocity V –a solution akin to a traveling wave solution” is typical and a whole body of literature exists, based on this restriction [26]. For the second aspects, the most important papers are reviewed in [25]. Clearly a velocity restriction does not allow the system to behave in a physically correct way and we risk getting flawed results.\\
\newline
The consequences are evidenced by the following example, taken from [37]. Fluid is pumped in a borehole at constant rate Q equal to 0.0001 m3/s causing the fracture to advance.\\
\newline
\begin{figure}[!htb]
   \begin{minipage}{0.5\textwidth}
     \centering
     \includegraphics[width=0.9\linewidth]{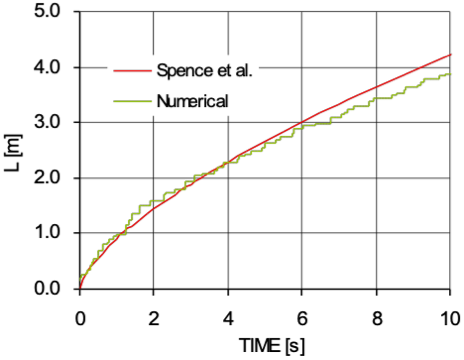}
   \end{minipage}\hfill
   \begin{minipage}{0.5\textwidth}
     \centering
     \includegraphics[width=0.9\linewidth]{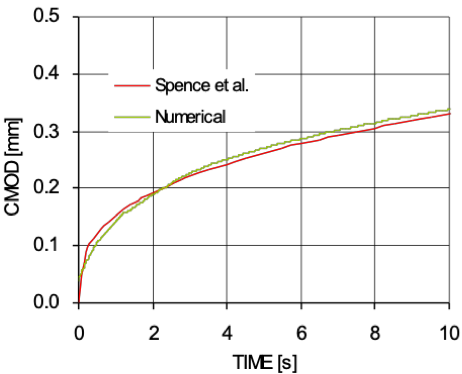}
   \end{minipage}
   \caption{Crack mouth opening displacement (on the right) and Crack length vs. time (on the left)}

\end{figure}
The following analytical solutions [38] for crack length L, crack mouth opening displacement (CMOD), and for the pressure at the crack mouth (Pcm) respectively comes from an asymptotic solution and an approximation to it, based on some simplifying assumptions; in particular the fluid is incompressible, fracture impermeable, and linear elastic fracture propagation takes place. Further, the fluid completely occupies the fracture volume, hence affecting the tip velocity and excluding fluid lag in the tip region.

\[
L=0.65\bigg(\frac{GQ^3}{\mu(1-\nu)}\bigg)^{\frac{1}{6}}t^{\frac{2}{3}}
\]
\[
CMOD=2.14\bigg(\frac{\mu(1-\nu)Q^3}{G}\bigg)^{\frac{1}{6}}t^{\frac{1}{3}}
\]
\[
p_{cm}=1.97\bigg(\frac{G^3 Q \mu}{(1-\nu)^3 L^2}\bigg)^{\frac{1}{4}}+S
\]

For the numerical solution, we have here adapted the cohesive model [31] to approach as much as possible linear elastic fracture and adopted a finite domain. Further the axial symmetry of the problem is broken by introducing a notch from which the crack enucleates. All the other simplifying assumptions of [38, 39] have been taken into account, Table 1. The numerical and analytical solutions are presented in Fig. 4. The crack mouth opening displacement matches pretty well, but there are some discrepancies with the fracture advancement although not too serious. The salient point is the non-regular stepwise advancement in the numerical solution. This is due to the fact that we use a crack-tip advancement/ time stepping algorithm shown in section 4 which respects the self-organization of rupture i.e. it imposes no tip velocity restriction. The pressure makes the difference.\\
\begin{figure}[b!]
\centering
\includegraphics[width=0.7\textwidth]{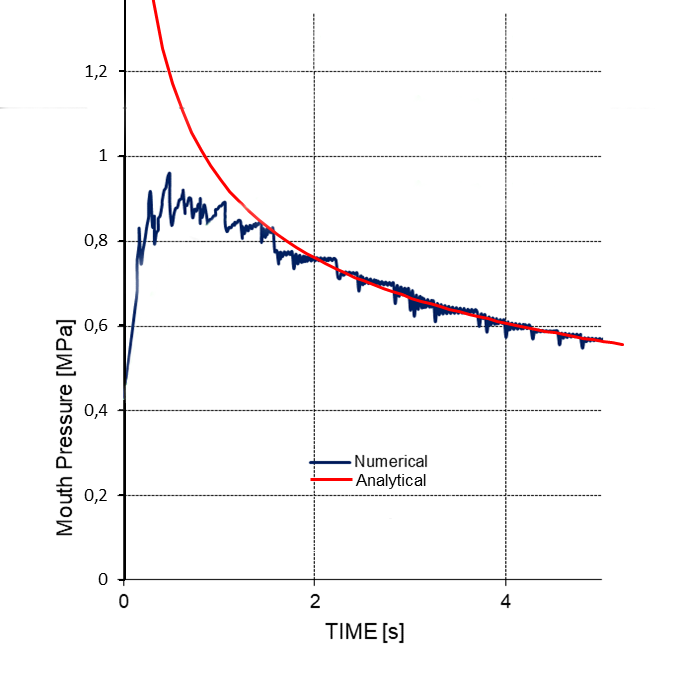}
\caption{Crack mouth pressure vs. Time (parameters used for the analytical solution, are shown in Table 1).}
\end{figure}
\begin{figure}[ht]
\centering
\includegraphics[width=0.8\textwidth]{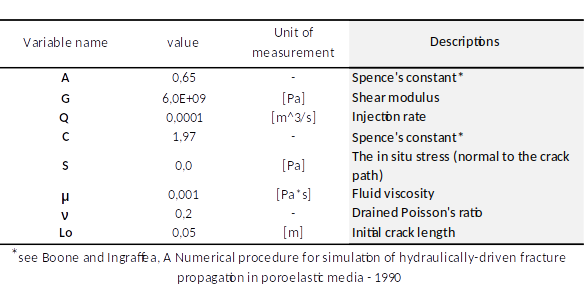}
\end{figure}
\begin{table}[ht]
\caption{Variables to be considered on the evaluation of interaction
  techniques}
\label{tab:exTable1}
\end{table}
After an initial period needed for starting, influencing the solution due to the initial condition chosen in the numerical model, the average pressure follows the exact solution, equation (3). While all analytical solutions [26] show smooth pressures versus time, in Figure 5 clearly downward pressure jumps are observed at each step of the fracture advancement which have been explained by Biot’s theory (second partial scenario of [27]): if flow is specified in the fluid continuity equation (second Biot equation, next section) the flow effect is transmitted to the solid through the pressure coupling term in the effective stress (first Biot equation). The solid is loaded and upon rupture produces a sudden increase of the volumetric strain and an ensuing drop in pressure. In quasi-static situations fluid lag does not matter much in this reasoning because it has been found [40] that in most practical cases fluid lags, if it exists at early stages, diminishes with the propagating fracture and may be ignored. In dynamics however, differences between the velocities of crack tip and fluid front in the fracture may be expected and may hence influence the phenomenon described. These pressure drops are observed both in experiments and in the field and make the solution relevant for steering fracking operation [28], and for investigating volcanic [15] and subduction tremor in geophysics [13]. In such situations, smooth solutions are perfectly useless and unphysical. In the next two sections we will show that these pressure jumps persist also in dynamic solutions.

\section{Dynamics in heterogeneous porous media}

The central force model previously developed for quasi-static situations [27] will be briefly recalled in the next sub-section and extended to consider inertia forces. In the same section, we will also explain how the concept of the self-organization of rupture is built inside this model. In sub-section 3.2 we will present two applications. The first one with a central flux imposed (hydraulic loading), the second one with a pressure imposed at the centre (mechanical loading). These examples which respect the requirement ii) of section 2 show that:
\begin{itemize}
	\item there are still the pressure rises and drops at fracture, as in the quasi static case;
	\item accounting for inertia will lead to a more physical description of the pressure oscillations and the fracture advancement.
\end{itemize}

\subsection{The model}
The model is the continuous-damage fully saturated model (CDFSM) of [27] based on the generalized Biot’s theory [41,42] under the following assumptions:
\begin{itemize}
	\item linear elastic stress–strain relations;
	\item small strains;
	\item incompressible solid and fluid phases. 
\end{itemize}
The extension of the CDFSM to dynamics is here summarized. The velocity of the fluid $\boldsymbol{v}^w$ is defined taking the velocity of the solid skeleton $\boldsymbol{v}^s$ as a reference and introducing the relative velocity $\boldsymbol{v}^{ws}$
\begin{equation}
\boldsymbol{v}^{ws}=\boldsymbol{v}^w-\boldsymbol{v}^s
\end{equation}
The water acceleration $a^{w}$ will be [43]:
\begin{equation}
\boldsymbol{a}^w=\boldsymbol{a}^s+\frac{D^s\boldsymbol{v}^{ws}}{Dt}+\grad{(\boldsymbol{v}^s+\boldsymbol{v}^{ws})}\cdot \boldsymbol{v}^{ws}=\boldsymbol{a}^s+\boldsymbol{a}^{ws}+\grad{(\boldsymbol{v}^s+\boldsymbol{v}^{ws})}\cdot \boldsymbol{v}^{ws}
\end{equation}
The last term in (2) is called convective term and $a^{ws}$ is the acceleration of the fluid with respect to the water. Following Zienkiewicz et al. [44], those terms are not taken into account. With these assumptions the linear momentum balance equation is as follows:
\begin{equation}
-\rho \boldsymbol{a}^s +\rho \boldsymbol{g} +\nabla \cdot \boldsymbol{\sigma} =0
\end{equation}
where $\rho^s$,$\rho^w$,$\rho$ are respectively the intrinsic density (ID) of the solid phase, the ID of water and $\rho=(1-n)\rho^s+n\rho^w$ is the averaged density of the multiphase systems. $n$ is the porosity. $\boldsymbol{a}^s$ is the acceleration of the solid matrix. $\nabla$ is the divergence operator and $\boldsymbol{\sigma}$ the total stress tensor; $\boldsymbol{g}$ is the acceleration related to gravitational effects.
Terzaghi’s effective stress principle holds and it is written in the form:
\begin{equation}
\boldsymbol{\sigma}=\boldsymbol{\sigma}'-\delta p^w
\end{equation}
with $\boldsymbol{\sigma}'$ the effective stress tensor, $p^w$ the pressure and $\delta$ the Kronecker delta. The continuity of the solid and fluid phases, together with Darcy’s law for fluid flow in a porous medium, yield:
\begin{equation}
\frac{\partial{\boldsymbol{\varepsilon}}}{\partial{t}}+\divv{\bigg[\frac{\boldsymbol{k}}{\mu^w}(-\grad{p^w}+\rho^w(\boldsymbol{g}-\boldsymbol{a}^w)) \bigg]}=0
\end{equation}
where $\boldsymbol{\varepsilon}$ is the volumetric strain, $\boldsymbol{k}$ and $\mu^w$ are the fluid intrinsic permeability and dynamic viscosity respectively (both assumed constant in space and time). Considering the water acceleration in the water mass balance equation will lead to a so called “dynamic seepage contribution”. Its effect is negligible, [44], and will not be considered.\\
Discretizing Eqs. (3) and (5) by means of the Galerkin Finite Element method, the coupled system of equations assumes the same form as in [27] except for the equilibrium equation of the mixture solid plus fluid (first equation) that now accounts for inertia term (6). The second equation is the continuity equation of the fluid (7).
\begin{equation}
\int_{\Omega}^{}\boldsymbol{B}^T \boldsymbol{\sigma}' d\Omega-\boldsymbol{Q} \bar{p}_w +\boldsymbol{M} \ddot{\bar{u}}=f_u
\end{equation}
\begin{equation}
\boldsymbol{H} \bar{p}_w +\boldsymbol{Q}^T \dot{\bar{u}}=f_p
\end{equation}
$\boldsymbol{M}$ is the mass matrix,$\boldsymbol{B}$  is the strain operator,$\boldsymbol{Q}$  the coupling matrix,$\boldsymbol{H}$  the permeability matrix, $\boldsymbol{\sigma}'$  the effective stress tensor responsible for deformation and strength,$\bar{p}_w$  and $\bar{u}$ respectively the nodal values of the pressure and of the displacements and $f_u$ and $f_p$ the loading terms. Dot means differentiation with respect to time.
$\int_{\Omega}^{}\boldsymbol{B}^T \boldsymbol{\sigma}' d\Omega=\boldsymbol{K}\bar{u}$, $\boldsymbol{M}\ddot{\bar{u}}$ and the other matrices have been derived for the three-layer element used in the following and are shown in Figure 6.\\
\begin{figure}[ht]
\centering
\includegraphics[width=1\textwidth]{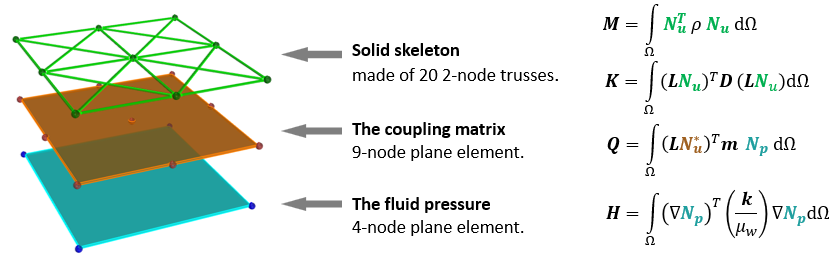}
\caption{Representation of the three-layer unitary element of the model (left) and the related matrices. The solid is represented by a square lattice of 20 trusses featuring vertical and horizontal bonds, together with diagonal ones; the related shape function (SF) is expressed by the symbol $\boldsymbol{N_u}$. The coupling matrix Q is obtained from a 9-node plane finite element (SF$\boldsymbol{N_u^*}$) and the flow H matrix from a 4-node element (SF$\boldsymbol{N_p}$). For more details see [27].}
\end{figure}
The system of equations (6) and (7) can be conveniently rewritten as:
\begin{align}
\begin{gathered}
\begin{bmatrix}
    \boldsymbol{M} & 0 \\
          0        & 0
\end{bmatrix}
\begin{bmatrix}
    \ddot{\bar u} \\
    \ddot{\bar p}_w
\end{bmatrix}
+
\begin{bmatrix}
          0          & 0 \\
    \boldsymbol{Q}^T & 0
\end{bmatrix}
\begin{bmatrix}
    \dot{\bar u} \\
    \dot{\bar p}_w
\end{bmatrix}
+
\begin{bmatrix}
  \boldsymbol{K}  & -\boldsymbol{Q} \\
         0        &  \boldsymbol{H}
\end{bmatrix}
\begin{bmatrix}
    \bar u \\
    \bar p_w
\end{bmatrix}
=
\begin{bmatrix}
    f_u \\
    f_p
\end{bmatrix}
\\
\downarrow
\\
 \boldsymbol{M} \ddot{\boldsymbol{a}}+\boldsymbol{C} \dot{\boldsymbol{a}} +\boldsymbol{K} \boldsymbol{a}= \boldsymbol{f} 
\end{gathered}
\end{align}
The time integration of the system (8) is carried out by the “Truncated Taylor series collocation algorithm GN22” [45]. The solution for the unknowns $\boldsymbol{a}=\begin{bmatrix} \bar u & \bar{p}_w\end{bmatrix}$ at a generic time step $t_{n+1}=t_n+\Delta t$ is given by:
\begin{align}
\begin{gathered}
\boldsymbol{a}_{n+1}=-\boldsymbol{A}^{-1}(\boldsymbol{f}_{n+1}+\boldsymbol{C}\dot{\hat{\boldsymbol{a}}}_{n+1}+\boldsymbol{M}\ddot{\hat{\boldsymbol{a}}}_{n+1})   
\\
\boldsymbol{A}=\frac{2}{\beta_2\Delta{t}^2}\boldsymbol{M}+\frac{2\beta_1}{\beta_2\Delta{t}}\boldsymbol{C}+\boldsymbol{K}
\\
\dot{\hat{\boldsymbol{a}}}_{n+1}=-\frac{2\beta_1}{\beta_2\Delta{t}}\boldsymbol{a}_n+\bigg(1-\frac{2\beta_1}{\beta_2}\bigg)\dot{\boldsymbol{a}}_{n}+\bigg(1-\frac{\beta_1}{\beta_2}\bigg)\Delta{t}\ddot{\boldsymbol{a}}_{n}
\\
\ddot{\hat{\boldsymbol{a}}}_{n+1}=-\frac{2}{\beta_2\Delta{t}^2}\boldsymbol{a}_n-\frac{2}{\beta_2\Delta{t}}\dot{\boldsymbol{a}}_{n}-\frac{1-\beta_2}{\beta_2}\ddot{\boldsymbol{a}}_{n}
\end{gathered}
\end{align}
where $\beta_2 \geq \beta_1 \geq 0.5$ are parameters for time integration.\\
In the CDFSM continuous damage is applied to the truss elements as follows. At the beginning of the simulation a uniform Young’s modulus of $100 MPa$ is assigned to each element. The spatial distribution of defect in the solid matrix (heterogeneous behavior) is taken in to account by randomly extracting a stress threshold from the range $(0,1) MPa$. Each threshold has the same probability of being extracted (uniform distribution). During the simulation, when the stress in a truss exceeds the local threshold, the elastic modulus of the truss is reduced by the factor $(1-D)$ , thus considering damage. $D$ is the fixed damage amount. The first extreme value $D = 0$ means no damage and the other extreme value $D = 1$ means that the truss is completely broken and does no more contribute to the global stiffness. To represent gradual damage, $D$ is set to $0.1$: each time that a truss is damaged, its new Young’s modulus value is $90\%$ of the previous one. Each truss can be damaged up to thirty times before the final failure, thus conveying an asymptotically decreasing damage rule. The threshold is updated after each damage event (annealed behavior).\\
No limits are imposed on the number of elements that could be damaged within a time step. The next time step starts only when, taken an element $i$ , the stress $\boldsymbol{\sigma}_i$ is less than the threshold $\boldsymbol{\sigma}^{max}_i$ for each $i$ , see Figure 7.\\
\begin{figure}[t]
\centering
\includegraphics[width=0.98\textwidth]{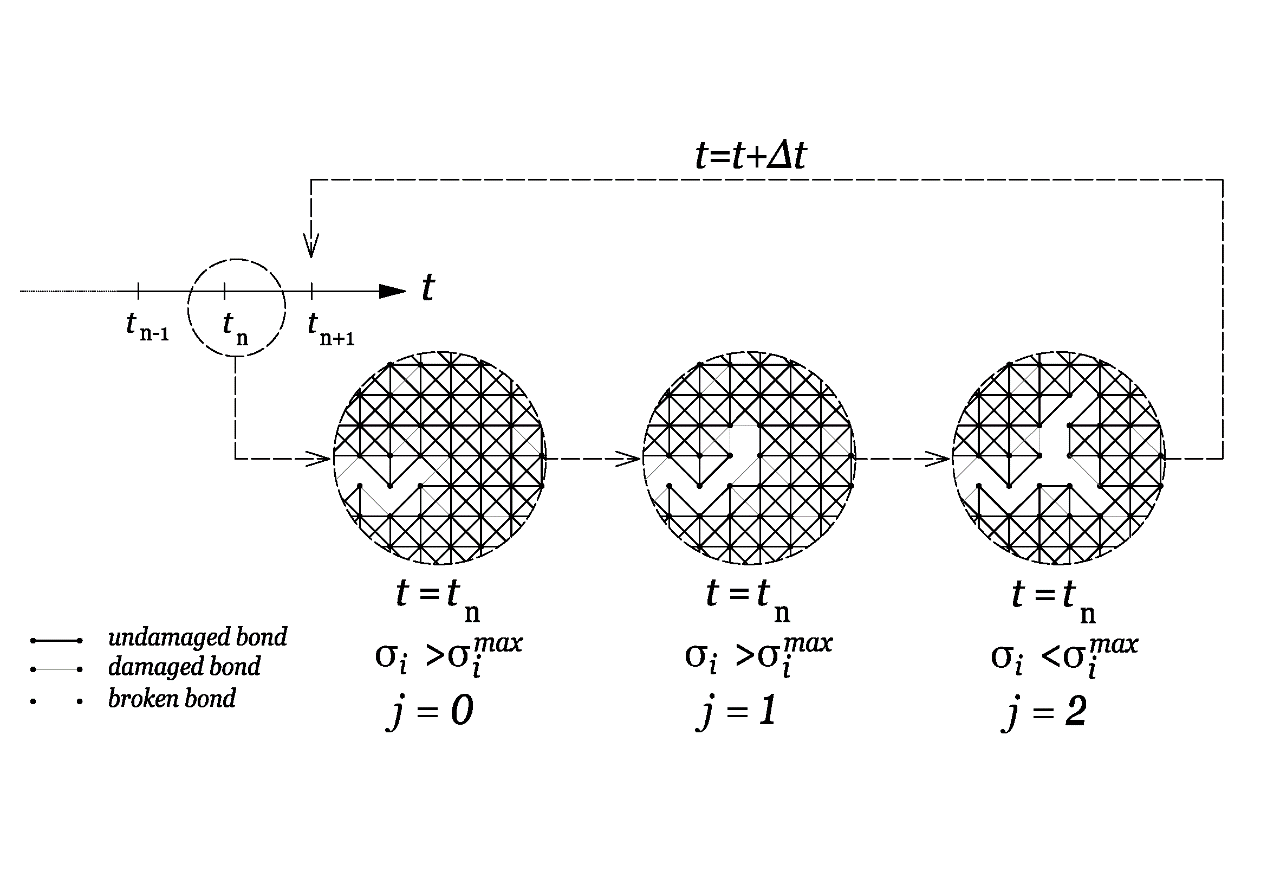}
\caption{Multiple advancing fracture steps at the same time station. The time step $\boldsymbol{\Delta t}$ refers to the loading time scale while $\boldsymbol j$ reefers to the internal loops related to SOC. This algorithm respects the self-organization of rupture.}
\end{figure}
The concept of the self-organization of rupture at the beginning of the section 2 is expressed in dynamics only by the requirement ii); while the requirement i) is implicitly satisfied. 
\subsection{Numerical applications}
The two examples of this section have the same domain, boundary and initial condition shown in Figure 8. The examples differ from each other because of the central boundary condition: a constant pressure is imposed $p(t)=1MPa$ in the example of 3.2.1, while a constant flux $f(t)=3*10^{-2}mm^3/s$ is imposed in the example 3.2.2.
\begin{figure}[ht]
\centering
\includegraphics[width=0.8\textwidth]{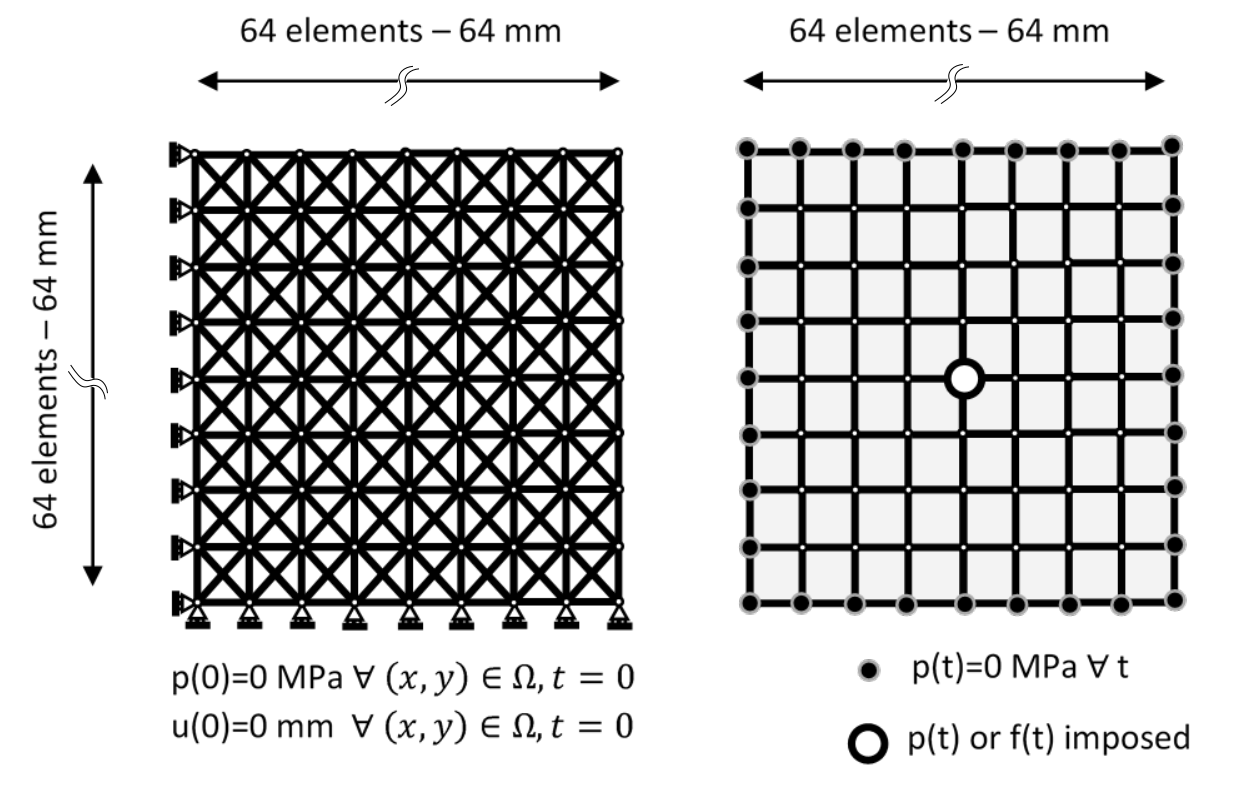}
\caption{Sketch of the domain consisting in total of $\boldsymbol{64}$ x $\boldsymbol{64}$ square elements of the type of Figure 6. Initial conditions and boundary conditions for the solid skeleton are shown in the left part; boundary conditions for the fluid in the right part of the figure. The additional parameters of the simulation are: Young modulus $\boldsymbol{E=100 MPa}$, Hydraulic Conductivity $\boldsymbol{K=0.001 mm^3/s}$.}
\end{figure}
Physically one can think about those examples as a plane strain representation of a soil specimen where a fluid is either pumped with a constant and a variable fluid flow in the centre. No mechanical load is applied. In both cases the permeability is kept constant, either in the porous matrix and in the fracture path. 
\subsubsection{Water pressure imposed at the centre of the domain}
This loading case is of mechanical type and is applied on the first Biot equation (8). In Figure 9, continuous line, both pressure rises and drops can be observed. This corresponds to the full scenario of [27], i.e. the second partial scenario of section 2 has to be completed by a first partial scenario which reads “if a load, pressure, or displacement boundary condition is applied suddenly (all these conditions acting on the equilibrium of the solid-liquid mixture), then the fluid bears initially almost all the induced load because its immediate response is undrained (rigid and non-flowing). Then through the coupling with the fluid, the overpressures decrease and the solid gets loaded”. This situation produces a pressure rise upon rupture. However, the sudden changes of displacements would induce also some pressure drop in this first partial scenario due to the volumetric strain and pressure rise in the second partial scenario; therefore, the effects of both scenarios will appear in general (whole scenario). In quasi-static case the two scenarios are well separated for mode I fracturing. In dynamics, this is not the case. In fact, when the sample is locally subjected to a sudden damaging process, or to a fracture advancement, part of the strain energy is released and converted to kinetic energy. Strain and displacement waves of finite velocity appear. During the simulation, we observe damage at multiple points, so we have multiple generation of strain waves the interaction of which can create conditions for new damage or new fracture advancement. This is influencing also the interaction between fluid and solid. The presence of these waves changes the behavior of the sample. Neglecting continuous damage would lead to a fast consolidation problem shown in Figure 9 with dashed lines while the continuous one represents the solution with continuous damage. Figure 10 shows the time evolution of the system at different time stations in terms of Young modulus (above) and stresses (below).
\begin{figure}[ht]
\centering
\includegraphics[width=0.8\textwidth]{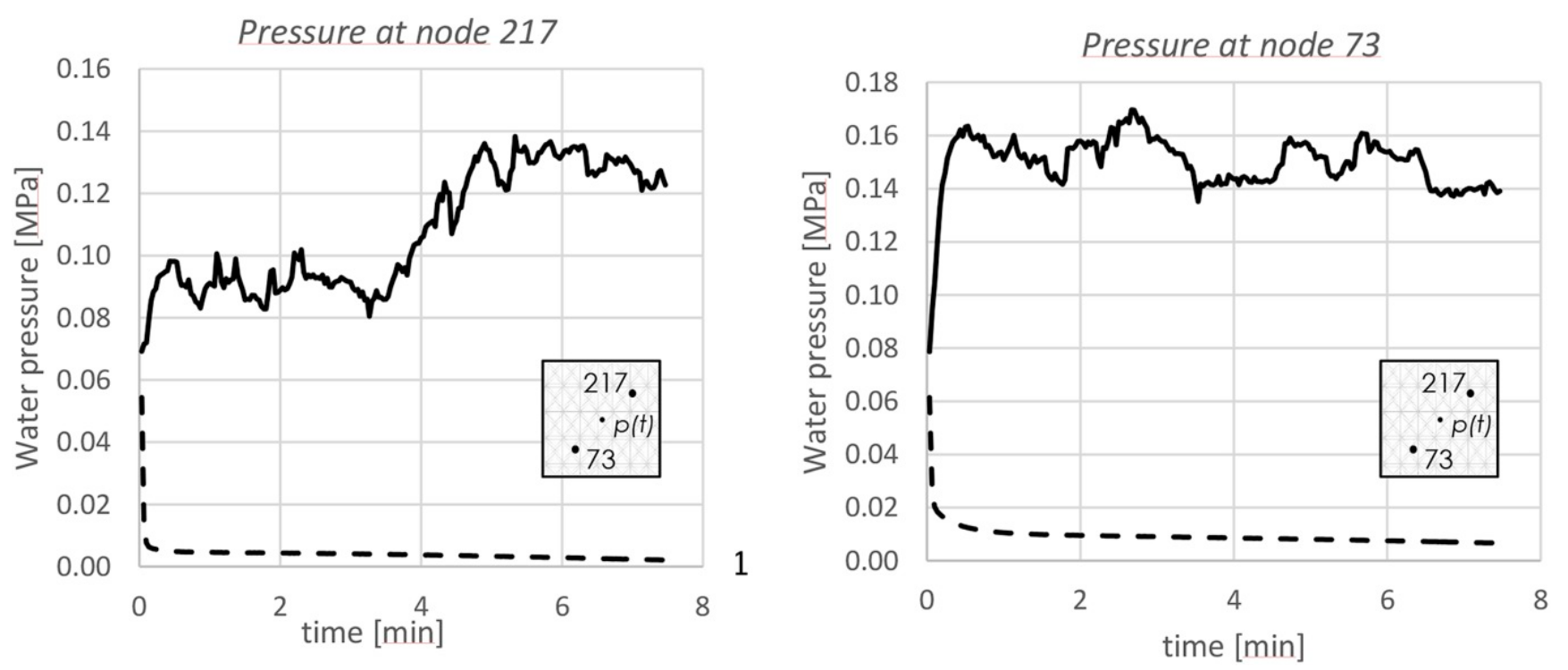}
\caption{Water pressure vs. time at two different positions of the domain (indicated in the boxes). The dashed line represents the solution of the same problem at the same positions but without the fracture algorithm.}
\end{figure}
\begin{figure}[ht]
\centering
\includegraphics[width=0.8\textwidth]{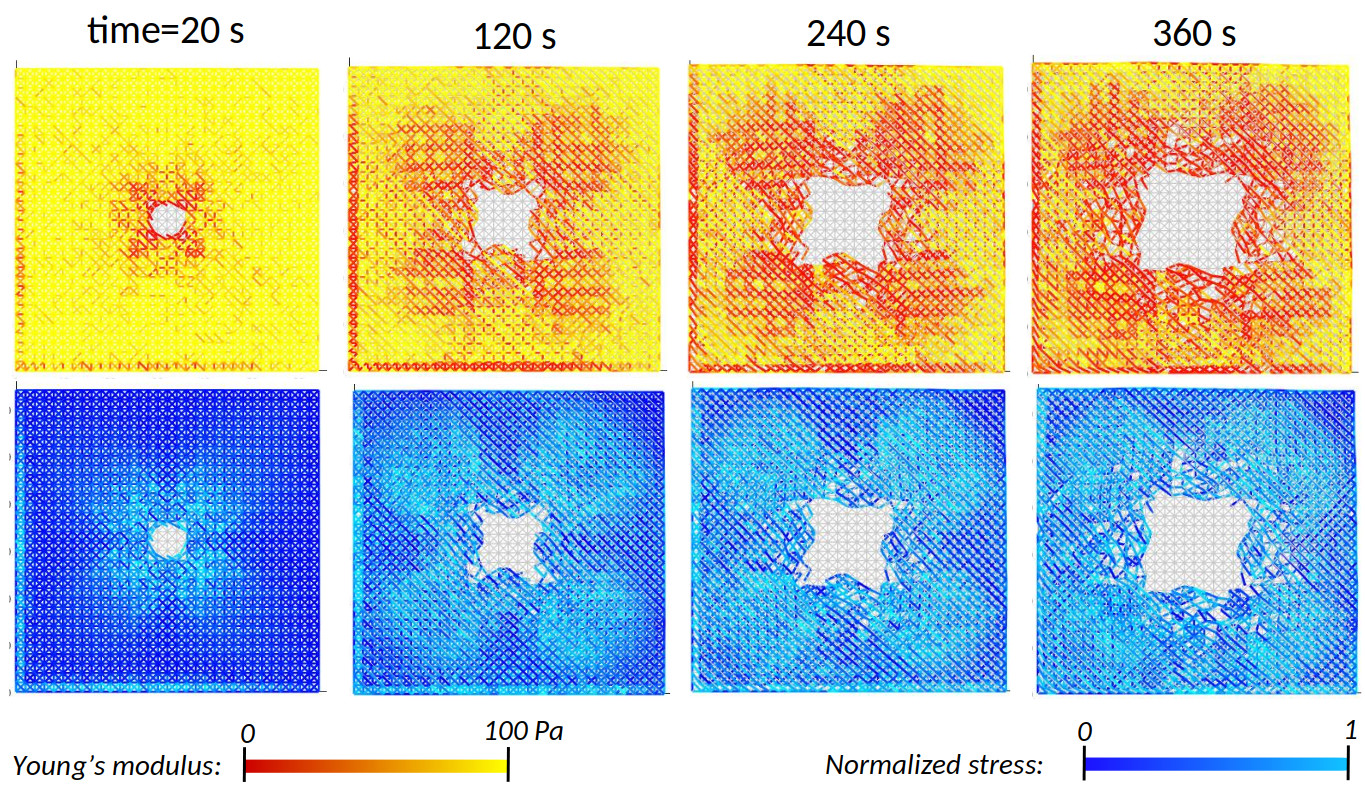}
\caption{From left to right the time evolution of the system is described at different time steps. For each of them the magnitude of the young modulus is represented (on the top row), while the absolute value of the stress is represented in the bottom row. The holes correspond to “broken” elements which have been eliminated.}
\end{figure}
\subsubsection{Imposed Flux}
In this second case a constant water flux is applied on the second Biot equation (8). Looking at Figure 11, we observe pressure both rises and drops: similarly to the previous section 3.2.1 we have the full scenario of [27]. The solution of the problem without the damage algorithm, dashed line in Figure 11, shows an increase of the pressure which slowly reaches a constant mean value (not shown in the Figure). Figure 12 shows the time evolution of the system at different time stations in terms of Young modulus (left) and stresses (right).\\
\begin{figure}[ht]
\centering
\includegraphics[width=0.8\textwidth]{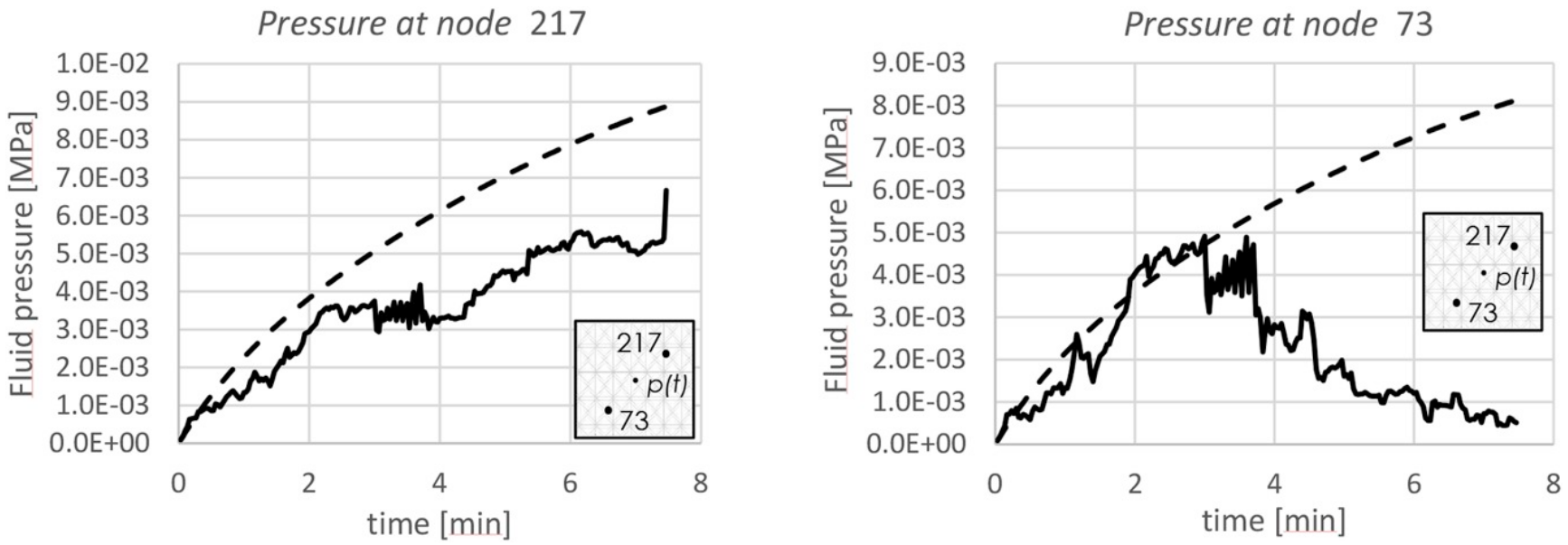}
\caption{Water pressure vs. time in two different positions in the domain (indicated in the boxes). The dashed line represents the solution of the same problem, in the same positions but without the fracture algorithm.}
\end{figure}
When the damage occurs at some position we mostly expect a pressure drop because of the new space created. This concept is proved in Figure 13, by comparing the state of the system before and after a pressure drop in the two pairs between steps 210 and 220 and steps 340 and 350. On the right-hand side, we see the evolution of pressure in some selected points (A-D), where the drops in the above intervals can be clearly seen.\\
Looking at the bottom pairs of pictures of Figure 13, the location of point 73 of Figure 11 and the related pressure evolution, we observe that at this node the overpressure is going to zero because of the nearby formation of a large fracture channel. On the contrary, at node 217 of Figure 11 (located in an un-fractured region, Fig. 13), the pressure is globally not decreasing even if it is still below the solution obtained without considering the fracture process. From Figure 12, we also note that on the boundary of an existing fracture, a new one has been nucleated and grows in a different direction from the previous one.\\
Note that the fracture width, unknown a priori, is bigger than the size of one cell of the lattice.\\
\begin{figure}[p]
\centering
\includegraphics[width=0.85\textwidth]{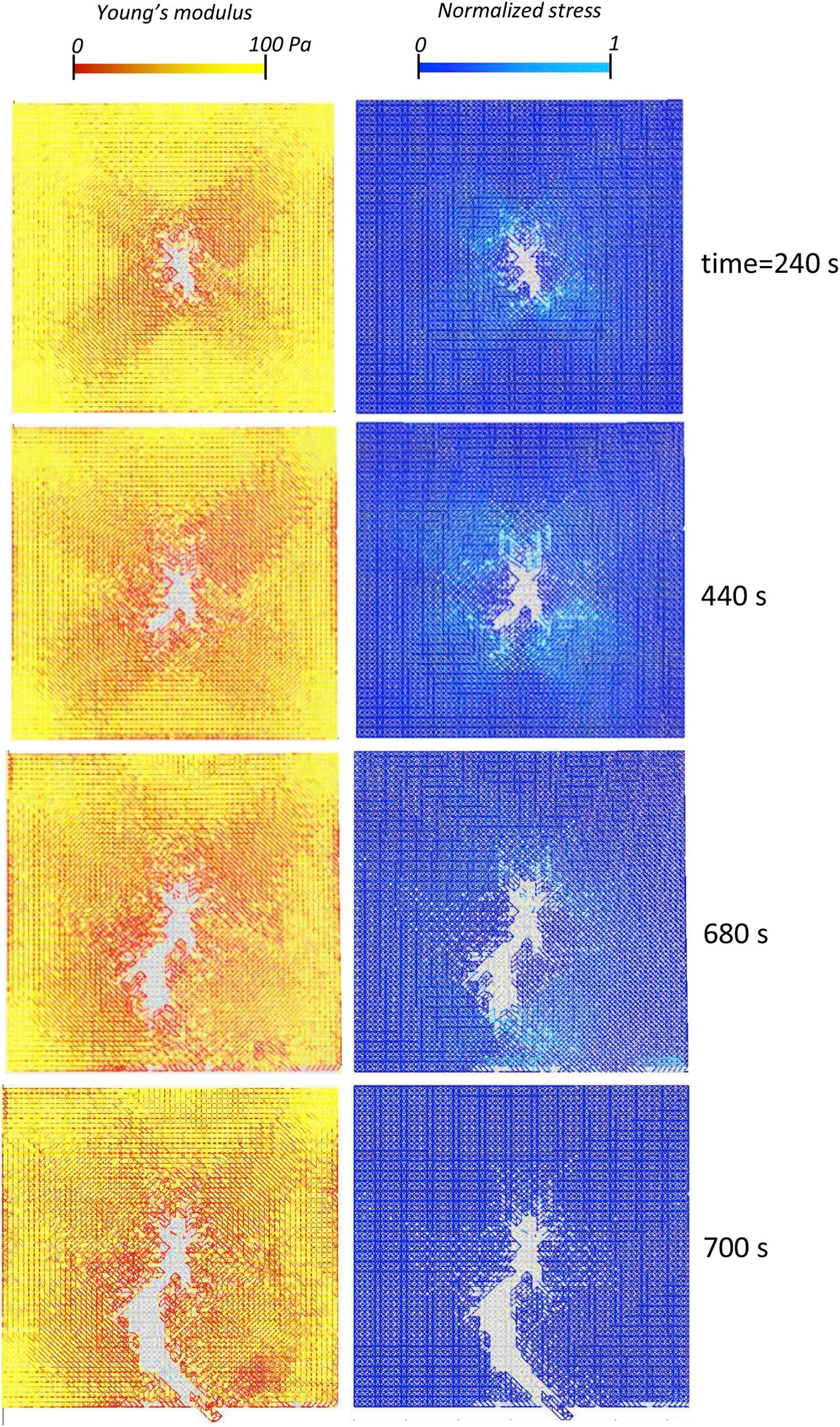}
\caption{From top to bottom: time evolution of the system at different time stations. For each of them the magnitude of the young modulus is represented on the left, while the absolute value of the stress is represented in the right column. The holes correspond to “broken” elements which have been eliminated.}
\end{figure}
\begin{figure}[p]
\centering
\includegraphics[width=0.9\textwidth]{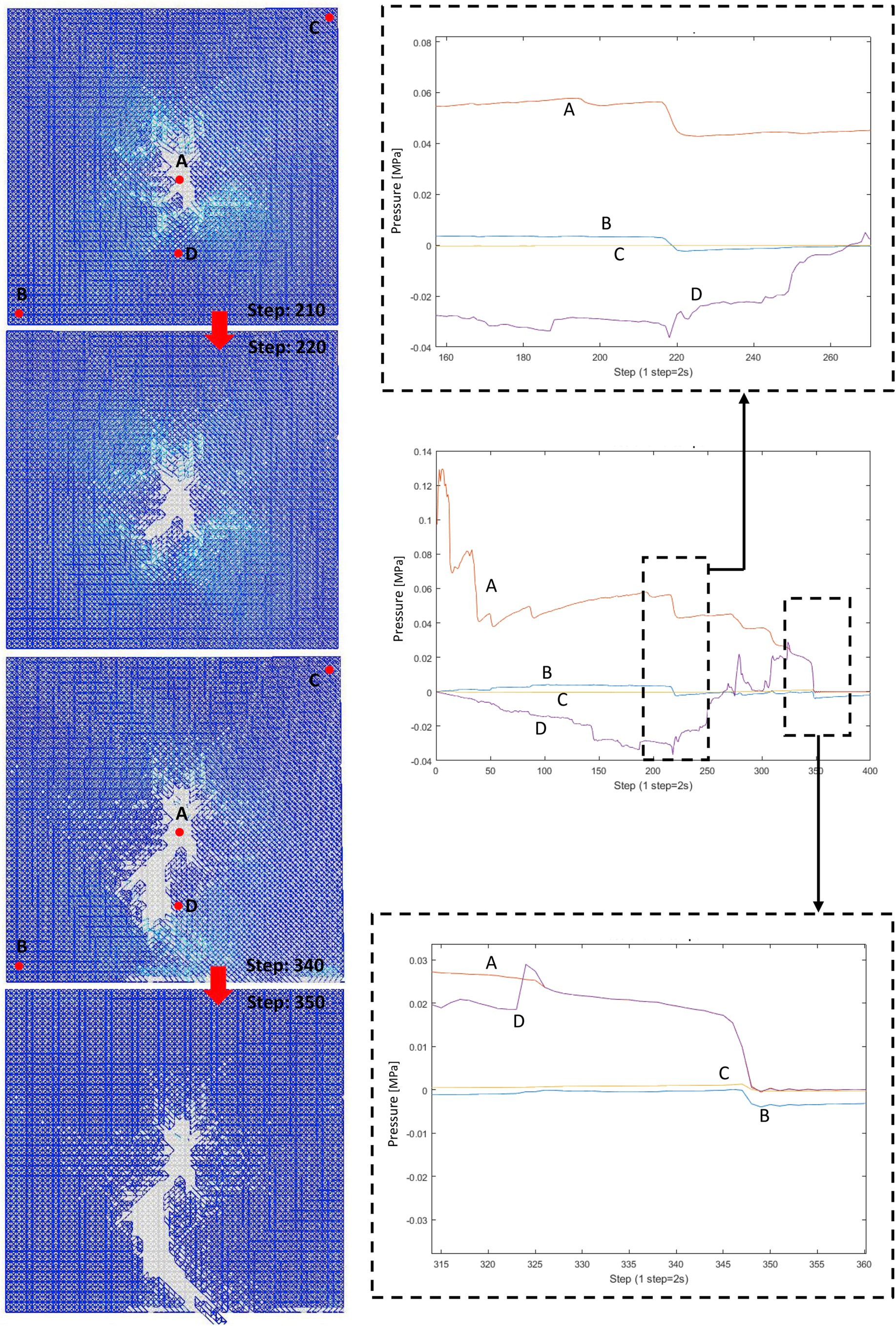}
\caption{Relation between the drops in the water pressure and the fracture advancement. On the centre of the right side of the figure, the behavior of the pressure is reported for different positions in the domain (points A-D). Below and above, two zooms of the central graph show the sudden pressure drop. A pair of images on the left upper part of the figure shows the domain immediately before and after the pressure drop recorded between steps 210 and 220. In the images, broken elements have been removed from the picture. A second pair of images on the right bottom part of the figure shows the domain immediately before and after the pressure drop recorded between steps 340 and 350.}
\end{figure}
\newpage
\section{Dynamics in homogeneous porous media}
In this section, we show a few results of two loading cases for fast dynamic fracturing in homogeneous saturated porous media taken from [25]. The following fracture advancement algorithm has been used [31] which respects the two requirements of section 2: a fracture arises when the Rankine criterion is not satisfied at a point. The fracture is then advanced by an increment $\Delta s$ in the direction normal to the maximum principal stress and the mesh is consequently re-built. 
\begin{figure}[ht]
\centering
\includegraphics[width=0.9\textwidth]{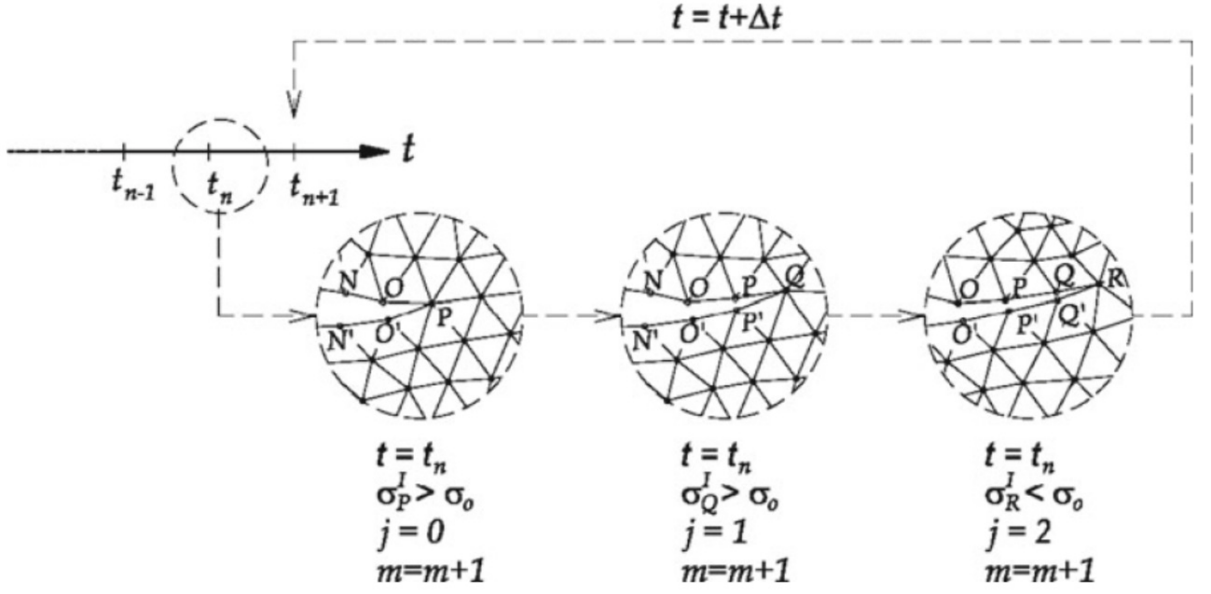}
\caption{Multiple advancing fracture steps at the same time station with ensuing Biot iterations. Redrawn with permission from Secchi et al. (2007).}
\end{figure}
In 2D the fracture follows directly the direction normal to the maximum principal stress while in 3D the fracture follows the face of the element around the fracture tip which is closest to the direction normal to the maximum principal stress; the fracture tip becomes a curve in space (front). If during the advancement a new node is created at the front the resulting elements for the fluid in the crack are triangular in 2D and tetrahedral in 3D. At each time $t_n$, $j$ successive tip (front) advancements are possible within the same time step; see Fig. 14. Their number in general depends on the chosen time step increment $\Delta t$, the adopted crack length increment $\Delta s$, and the variation of the applied loads. This algorithm works also in dynamics and in [25] it has been shown that still several advancements within a time step happen despite the fast loading. For another algorithm, which avoids remeshing but respects the two conditions of section 2, see [32].\\
he two loading cases are shown in Figure 15. The saturated porous medium sample is a rectangle of $0.2 m$ height and $2 m$ length. In case of mechanical loading the sample is loaded in traction by two uniform vertical velocities with magnitude $2.35$x$102 m/s$ applied in opposite directions to the left end of the top and bottom edges. In case of pressure loading the same cross section is subjected to a fast pressure increase of $5$x$10^{11} Pa/s$ at the crack mouth reaching a maximum value of $100 MPa$. Vertical and horizontal displacements are constrained at the right edge, and the boundary is impervious and adiabatic. For material parameters see [25]. In both cases stepwise advancement and pressure oscillations are obtained but the frequency content of these oscillations differ completely between the two loading cases: while for mechanical loading a broad spectrum appears with peaks at $2 kHz$, $12 kHz$ and $18 kHz$ , in case of hydraulic loading a single peak prevails, see also [13, 14].
\begin{figure}[t]
\centering
\includegraphics[width=0.88\textwidth]{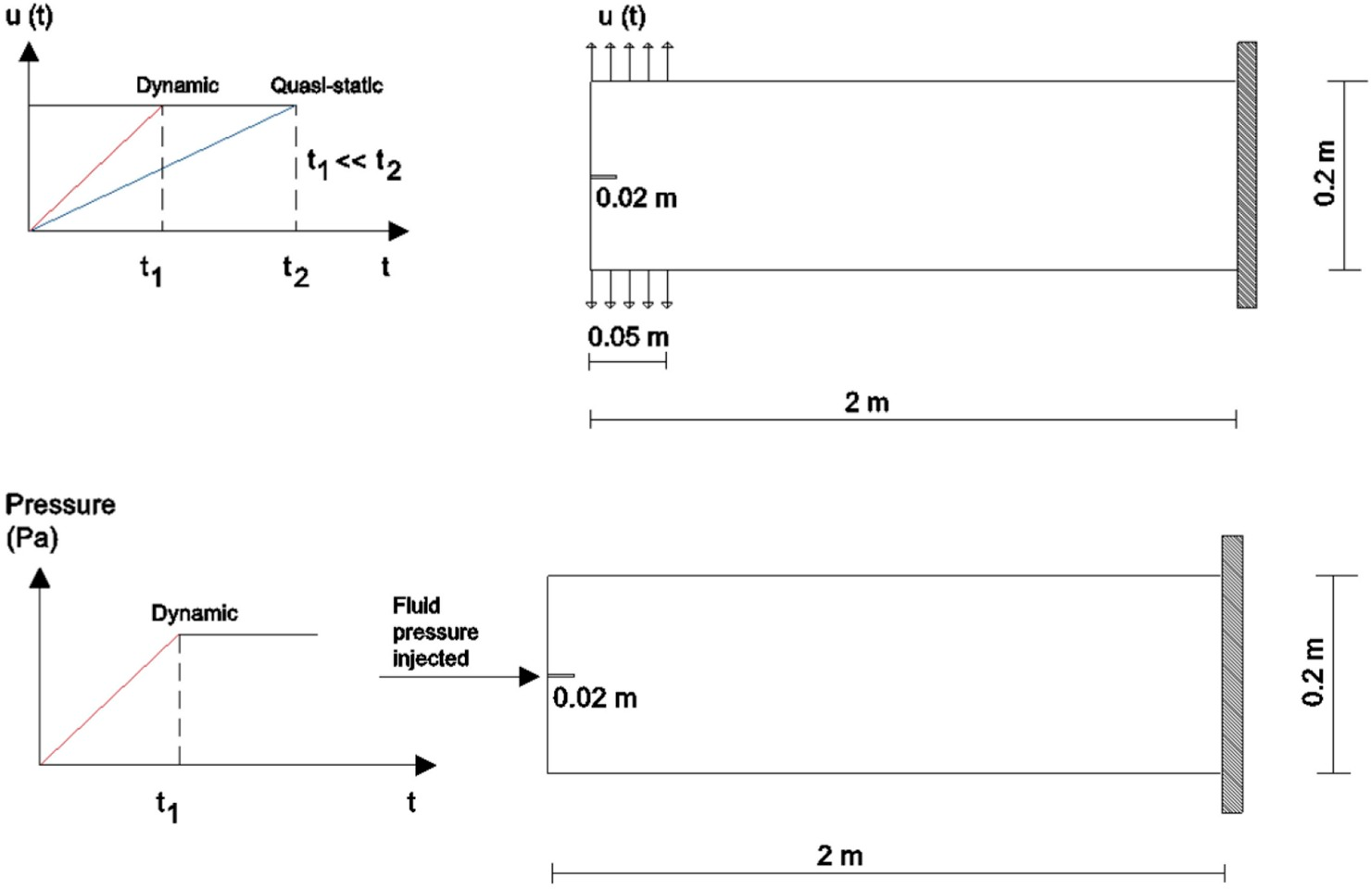}
\caption{Investigated cross section and loading cases. Redrawn with permission from [25].}
\end{figure}
We show in for mechanical loading Figure 16 a few significant snapshots for the pressure wave propagation at respectively (a) $0.02025 s$; (b) at $0.02035 s$; and (c) at $0.0205 s$.
\begin{figure}[b!]
\centering
\includegraphics[width=0.9\textwidth]{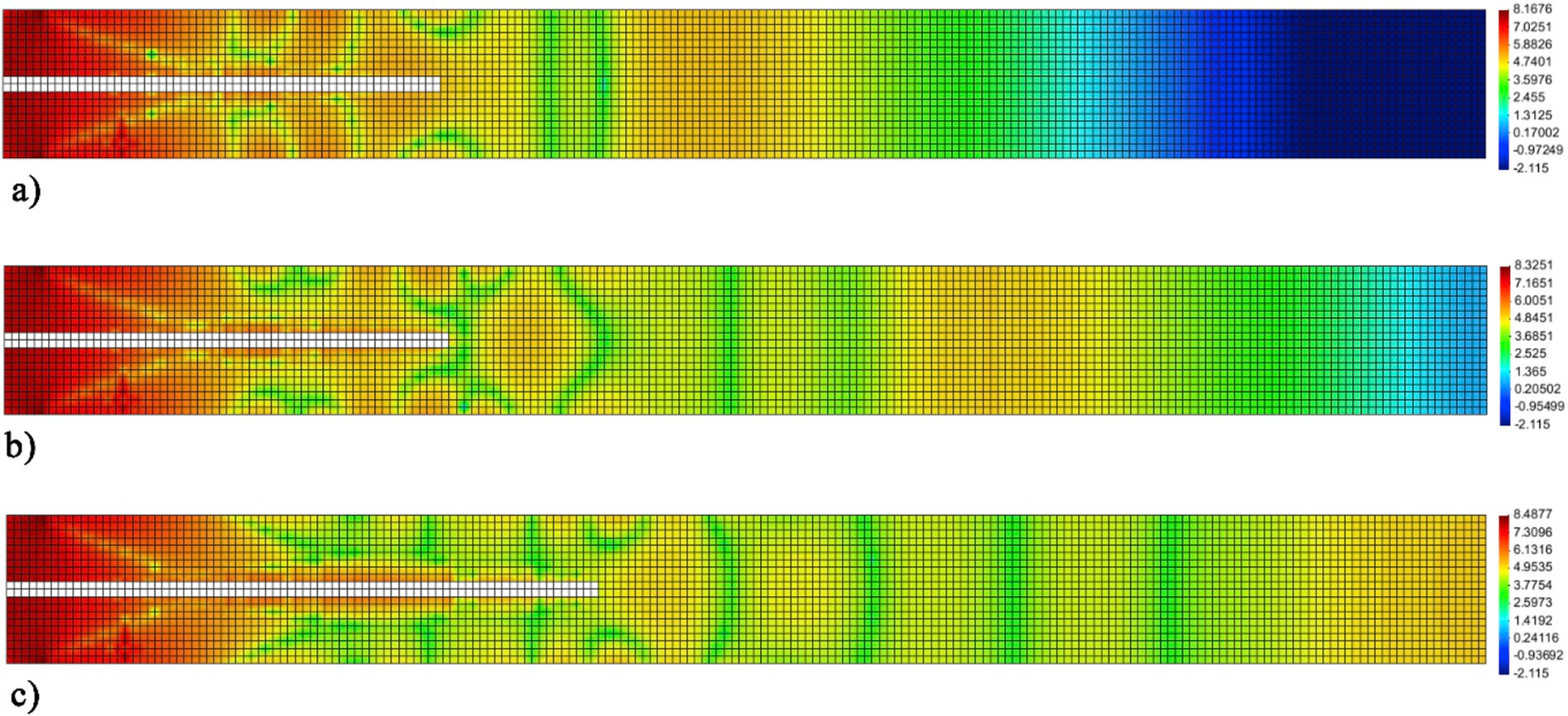}
\caption{Wave propagation of pressure contour plotted by logarithmic scale of GID software 12.4 at: (a) $\boldsymbol{0.02025 s}$; (b) $\boldsymbol{0.02035 s}$; and (c) $\boldsymbol{0.0205 s}$.}
\end{figure}
For pressure loading the pressure wave contours are depicted in Figure 17 at (a) $0.02005 s$; (b) $0.02025 s$; and (c) $0.0203 s$. Also from these graphs the different responses can be appreciated. Note that the chosen time instants differ from those of [25] and Figure 17 is much better resolved. Outgoing waves parallel to the crack can be noticed in this case.\\
\begin{figure}[ht]
\centering
\includegraphics[width=0.9\textwidth]{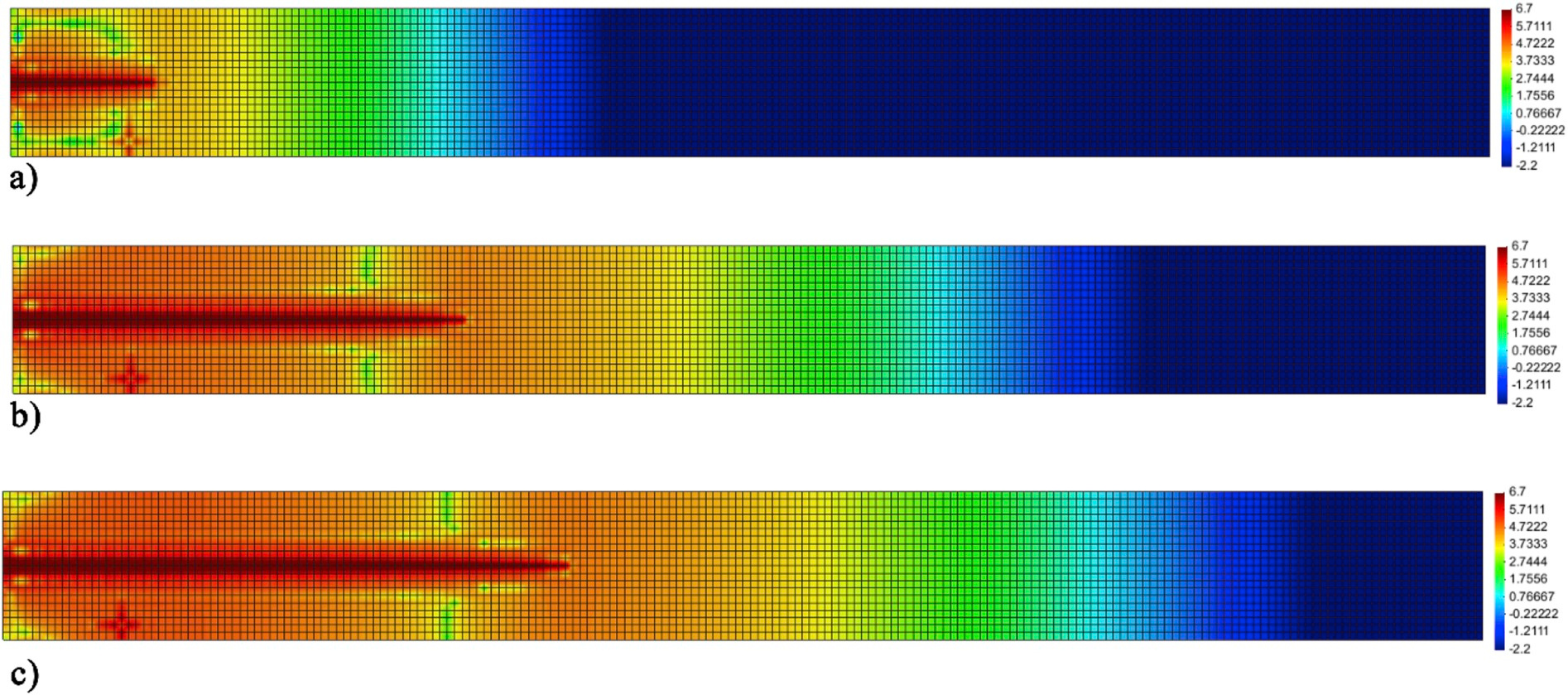}
\caption{Pressure wave contour plots of dynamic solutions for water pressure loading at (a) $\boldsymbol{0.02005 s}$; (b) $\boldsymbol{0.02025 s}$; and (c) $\boldsymbol{0.0203 s}$.}
\end{figure}
\section{Dynamic solution of a debonding beam on elastic foundation}
Fineberg et al. [46] experimentally studied instability of dynamic fracture on plastic polymethylmethacrylate samples (PMMA). These experiments reveal strong velocity oscillations. Tvergaard and Needleman [47] investigated numerically dynamic crack growth in a porous ductile material and found strong oscillations of J versus time at and after the fracture initiation. Avalanche behavior has been found in fracturing heterogeneous dry materials in quasi static conditions also [3, 27]. We investigate here dynamic fracturing on a dry material by adapting the model shown in Williams [48] which was used to assess the energy dissipated by a single debonding fiber in the cohesive zone using different cohesive forces laws. The mode is based on the deformation of a Euler-Bernoulli beam. We have extended the original model taking into account the inertia force and this first formulation is a precursor for a full model including also the interaction with the fluid phase. With this model, it is possible to show that the velocity of crack tip advancement is not constant but changes in time due to dynamics.\\
\begin{figure}[b]
\centering
\includegraphics[width=0.65\textwidth]{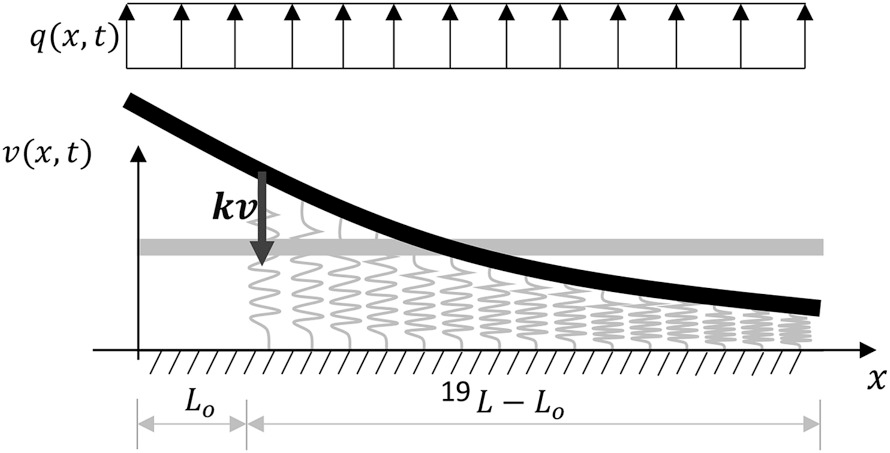}
\caption{Schematic representation of the model: initial (light grey) and deformed (black) configuration of the single fiber. The springs graphically represent the elastic bonds in the process zone.}
\end{figure}
We consider a fiber of total length $L$ as shown in Figure 18, bonded to a fixed support for a length $L-L_o$. The black line of Figure 18 represents the configuration of half of a crack of initial length $L_o$; the remaining bonded part, of length $L-L_o$ is the process zone and the fully bonded zone.\\
From the length $L_o$ to $L-L_o$, the behavior is described by the equation of a beam on an elastic foundation (10):\\
\begin{equation}
EJ\frac{\partial^4v(x,t)}{\partial{x^4}}+f(v)=q(x,t)-\rho A\frac{\partial^2v(x,t)}{\partial{t^2}}
\end{equation}
where $v(x,t)$ is the vertical displacement of the beam axis, i.e. the crack mouth opening (Figure 18), and it is a function of the spatial coordinate $x$ and the time $t$. $\rho A$ is the beam weight per unit of length and represents the inertia of the upper part of the solid body with regard to the fracture. The flexural rigidity $EJ$ of the beam represents the stiffness of the homogeneous medium above the fracture plane. $q(x,t)$ is the far field stress due to the external load applied to the body and in the following it is assumed constant in space (i.e. a uniformly distributed load is considered). Finally, $f(v)$ is the function expressing the cohesive forces. At the beginning of the process an initial fracture of length Lo is already formed, so there the cohesive forces are set to $0$ (i.e.$f(v)=0$ ) in equation (10). This leads to the following form of this equation which holds for the region between the origin $x= 0$ and $x=L_o$:\\
\begin{equation}
EJ\frac{\partial^4v(x,t)}{\partial{x^4}}=q(x,t)-\rho A\frac{\partial^2v(x,t)}{\partial{t^2}}
\end{equation}
Among the various formulations for the cohesive traction law proposed by Williams [48], we used the one shown in Figure 19 with three different values of threshold. Moreover, at the interface between the two regions proper continuity conditions should be assured. These are described by the following four equations defined at $x=x_{interface}$ for each time instant ($x_{interface} = L_o$ at the beginning):
\begin{equation}
\frac{\partial^iv^L(x_{interface},t)}{\partial{x^i}}=\frac{\partial^iv^R(x_{interface},t)}{\partial{x^i}}  , \; \;  i=0,...,3
\end{equation}
in which the superscripts $L$ and $R$ indicate the displacement $v$, respectively on the Left and on the Right side of the point at the interface. The coordinate $x_{interface}$ is changing in time considering that fracture is a moving boundary problem.\\
The final set of governing equations is formed by eqs. (10), (11) and (12) with the following boundary conditions defined on the left and right side respectively:\\
\begin{align}
\begin{gathered}
\frac{\partial^2v(0,t)}{\partial{x^2}}=0,\;\;\frac{\partial^3v(0,t)}{\partial{x^3}}=0 \; \; \forall t
\\
\frac{\partial^2v(L,t)}{\partial{x^2}}=0,\;\;\frac{\partial^3v(L,t)}{\partial{x^3}}=0 \; \; \forall t
\end{gathered}
\end{align}
which correspond to the free-end conditions. In this first application, we consider free boundary at the right-hand side which can produce wave reflection. In the next developments, we will consider   also a transmitting boundary. Note that here we have a fourth order non-homogeneous differential equation compared to a homogeneous second order one as usual in seismic analysis in a continuum [44].\\
The applied load is considered constant in space and increasing in time from zero up to a constant value. The external load is distributed for the whole length L of the system and it is directed according to the positive vertical axis, (Figure 18).\\
\begin{figure}[t]
\centering
\includegraphics[width=0.95\textwidth]{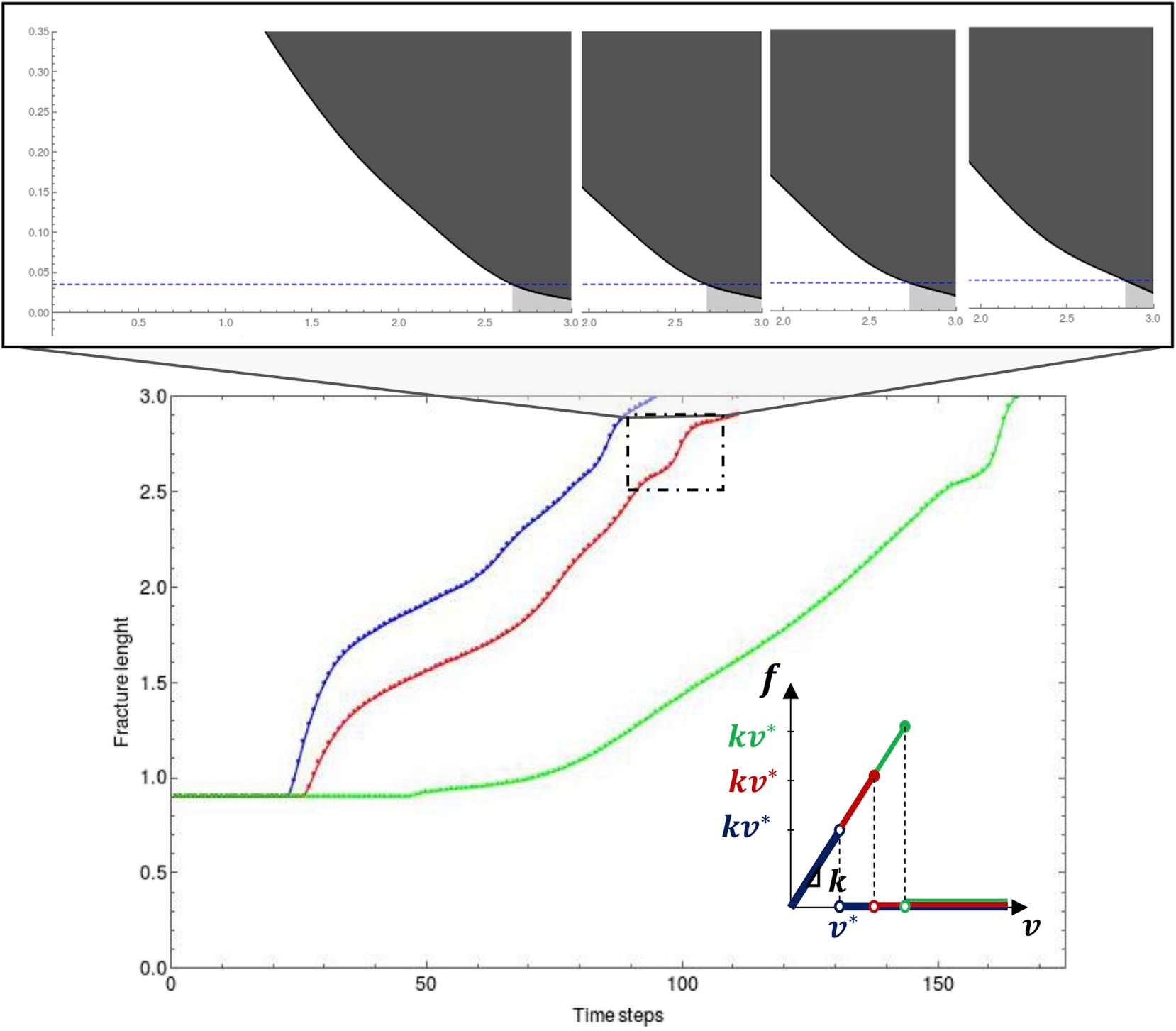}
\caption{Fracture length vs time for different thresholds in the force-opening constitutive law (bottom) and evolution of the cohesive zone for the intermediate case where change of the curvature can be observed (top).}
\end{figure}
tarting from an initial condition corresponding to zero velocity and zero displacement, we obtain the fracture length vs time depicted in Figure 19 for different thresholds in the constitutive law. It appears clearly that, due to the inertia forces, the fracture speed is not constant but alternates periods of fast growth with periods of slow down.\\
Figure 20 shows the evolution of crack opening during the simulation in three distinct time instants: at the beginning of the process for $x_{interf}=L_o$, (A), at an intermediate time station (B) and close to the complete debonding (C).
\begin{figure}[p]
\centering
\includegraphics[width=0.7\textwidth]{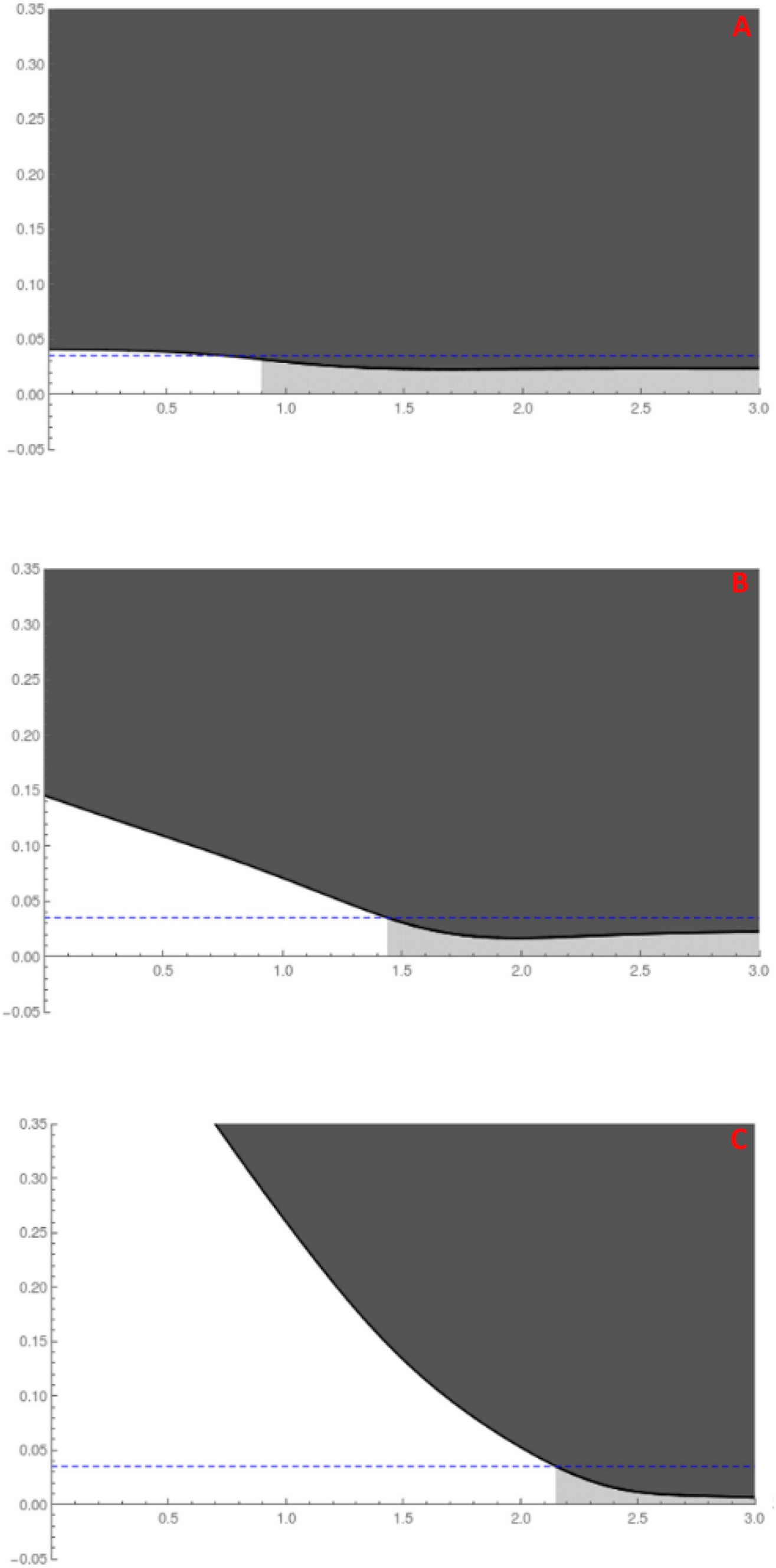}
\caption{Snapshots from A to C describe the evolution of the crack opening during the simulation. The light grey part represents the zone where the cohesive tractions are acting. The horizontal dashed line represents the maximum opening condition where we use the continuity equation (12).}
\end{figure}

\section{Conclusions}
By combining insight from statistical physics with that from fracture mechanics, both analytical and computational, a new light has been shed on simulation of fracture advancement. We have identified two requirements coming from statistical physics dealing with self-organization of rupture which should be included in the fracture advancement rule [2, 3]. These requirements are:  i) the external drive has a much slower timescale than fracture propagation; and ii) the increment of the external load (drive) is applied only when the internal rearrangement of fracture is over. Actually ii) is a consequence of i), and they constitute a "quasi-static" simulation. We show that the second requirement works also well in dynamics situations. These requirements should hence be taken into account when designing experiments and algorithms if we want to observe clean SOC. Ignoring them gives rise to conceptually defective algorithms yielding unphysical results. But experimentally and numerically speaking they are a choice: one can also load faster and then not be able to discern individual avalanches anymore. Most numerical algorithms and exact solutions do this and yield smooth results both for tip advancement and pressure, when carefully carried out experiments and field observations show the opposite. Finally, we show with a debonding beam on elastic foundation model that for a dry material the velocity of crack tip advancement is not constant but changes in time due to dynamics. In this model, the interaction with a fluid will be introduced. 

\section*{Acknowledgements}
The authors thank Professor H.J. Herrmann for most useful comments. The work of B.A. Schrefler, has been carried out with the support of the Technische Universität München - Institute for Advanced Study, funded by the German Excellence Initiative and the TÜV SÜD Foundation. C. Peruzzo and F. Pesavento were supported by the project 734370-BESTOFRAC (“Environmentally best practices and optimisation in hydraulic fracturing for shale gas/oil development”)-H2020-MSCA-RISE-2016.

\end{document}